\def \SAIT #1 #2 {{\em Mem.\ Soc.\ Astron.\ It.\/} {\bf #1}, #2}
\def \MESS #1 #2 {{\em The Messenger\/} {\bf #1}, #2}
\def \ASTRNACH #1 #2 {{\em Astron. Nach.\/} {\bf #1}, #2}
\def \AAP #1 #2 {{\em Astron. Astrophys.\/} {\bf #1}, #2}
\def \AAL #1 #2 {{\em Astron. Astrophys. Lett.\/} {\bf #1}, L#2}
\def \AAR #1 #2 {{\em Astron. Astrophys. Rev.\/} {\bf #1}, #2}
\def \AAS #1 #2 {{\em Astron. Astrophys. Suppl. Ser.\/} {\bf #1}, #2}
\def \AJ #1 #2 {{\em Astron. J.\/} {\bf #1}, #2}
\def \ANNREV #1 #2 {{\em Ann. Rev. Astron. Astrophys.\/} {\bf #1}, #2}
\def \APJ #1 #2 {{\em Astrophys. J.\/} {\bf #1}, #2}
\def \APJL #1 #2 {{\em Astrophys. J. Lett.\/} {\bf #1}, L#2}
\def \APJS #1 #2 {{\em Astrophys. J. Suppl.\/} {\bf #1}, #2}
\def \APSS #1 #2 {{\em Astrophys. Space Sci.\/} {\bf #1}, #2}
\def \ASR #1 #2 {{\em Adv. Space Res.\/} {\bf #1}, #2}
\def \BAIC #1 #2 {{\em Bull. Astron. Inst. Czechosl.\/} {\bf #1}, #2}
\def \JSQRT #1 #2 {{\em J. Quant. Spectrosc. Radiat. Transfer\/} {\bf #1}, #2}
\def \MN #1 #2 {{\em Mon. Not. R. Astr. Soc.\/} {\bf #1}, #2}
\def \MEM #1 #2 {{\em Mem. R. Astr. Soc.\/} {\bf #1}, #2}
\def \PLR #1 #2 {{\em Phys. Lett. Rev.\/} {\bf #1}, #2}
\def \PASJ #1 #2 {{\em Publ. Astron. Soc. Japan\/} {\bf #1}, #2}
\def \PASP #1 #2 {{\em Publ. Astr. Soc. Pacific\/} {\bf #1}, #2}
\def \NAT #1 #2 {{\em Nature\/} {\bf #1}, #2}

\documentstyle{memsait2}

\input epsf.sty
\begin{opening}
\title{THE MEASUREMENT OF THE SKY BRIGHTNESS AT MERATE OBSERVATORY}
\author{E. Poretti,  M. Scardia}
\institute{Osservatorio Astronomico di Brera, Milano, Italy}
\date{} 
\end{opening}

\begin{document}
\oddpagefooter{}{}{} 
\evenpagefooter{}{}{} 
\
\bigskip
\begin{abstract}

We describe the problems met using the telescopes of Merate Observatory, the
branch of Brera Astronomical Observatory.
From the point of view of the data analysis, we discuss how to introduce a
satisfactory correction of the variable extinction coefficient, giving also
some examples. We also measured the sky brightness, obtaining $V$=18.2 
mag/arcsec$^2$. The light pollution is responsible for such a  bright 
background and we present pictures showing its effects. We are limiting
it trying to persuade public administrations to reduce
the light scattered toward the sky by using cutoff lamps and putting
out commercial searchlights. The need of a law which safeguards 
astronomical activities is stressed.
\end{abstract}

\section{The telescopes of Merate Observatory}
Since a long time, the utility of small telescopes located near to research
Institutes is re--evaluated, as they can provide useful surveys of variable
stars and peculiar objects, especially over a long--time baseline. This kind of
survey cannot be further done at large observatories, owing to competition
in dividing time at big telecopes and the closing of the small
ones for budget problems.

For this reason, a constant effort was made to keep the Merate telescopes 
at a reasonable level of performances, in order to carry out such long--terms
programs, often requiring high--precision measurements too (photometry with a 
precision better than 0.010 mag, for example). The 50--cm Marcon telescope is
equipped with a photon--counting, single channel photometer; $UBV$ and $uvby\beta$
filters are available. Usually, stars brighter than 9.5 mag are measured with
the requested accuracy. For fainter objects,  we have equipped the 102--cm
Zeiss telescope with a CCD camera, allowing us to observe successfully 
fainter objects.  Moreover, we are now substituting the optics of the Ruths 
telescope, replacing the old 137--cm metallic mirror with a 134--cm Astrositall
one; to do that, the generous contribution of {\sc cariplo} Foundation was
of paramount importance. The substitution will be completed in late 1998; it
is foreseen to automatize the movements of the telescope. 

The three domes are also open to public to sightsee celestial objects in
some nights each month, usually around the Moon first quarter.
\begin{figure}
\epsfysize=19truecm 
\epsfxsize=13truecm 
\hspace{3.5cm}\epsfbox{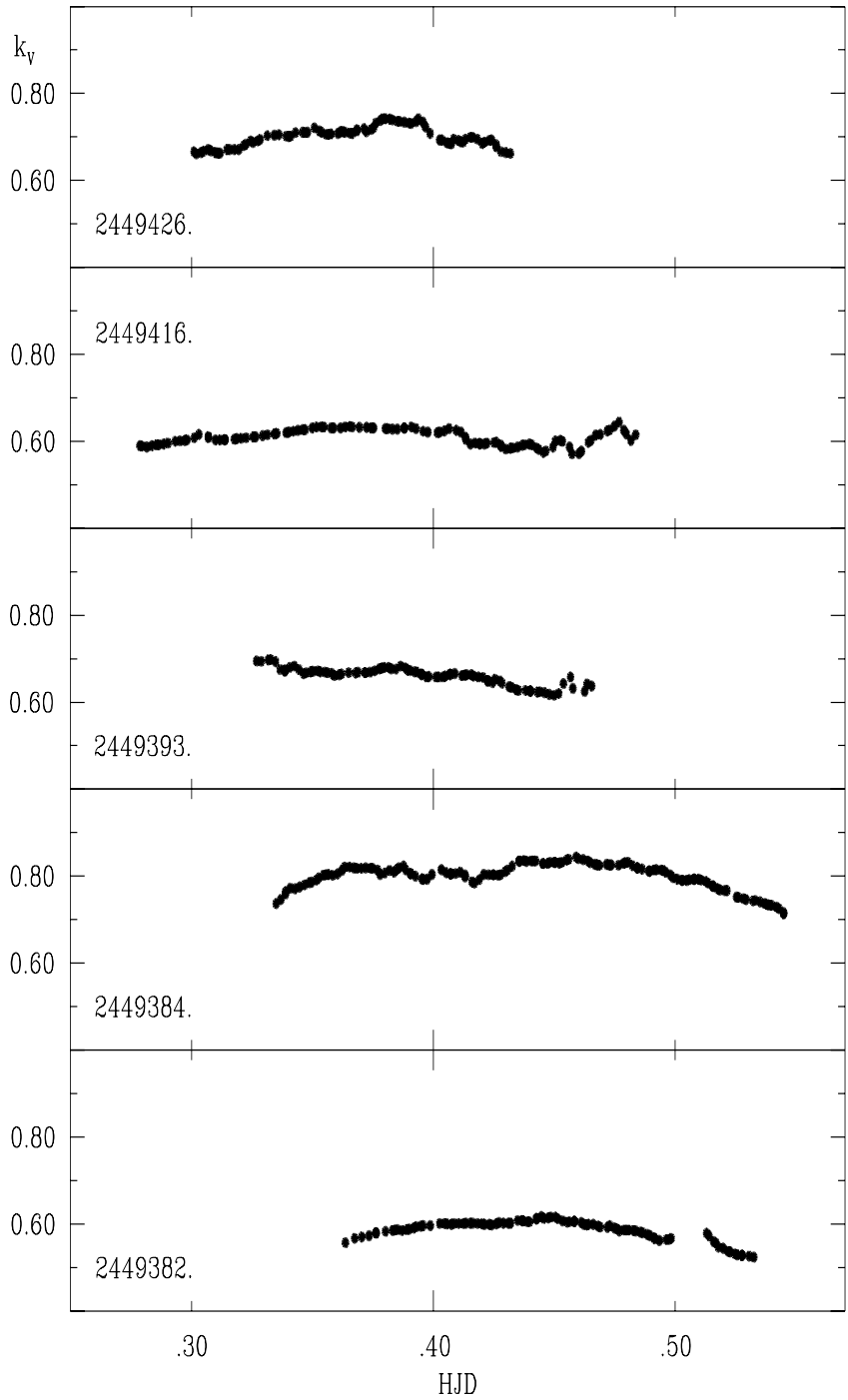} 
\caption[h]{Behaviour of the extinction coefficient in $V$ light as observed
at Merate Observatory on some nights in early 1994.}
\end{figure}

\begin{figure}
\epsfysize=19truecm 
\epsfxsize=13truecm 
\hspace{3.5cm}\epsfbox{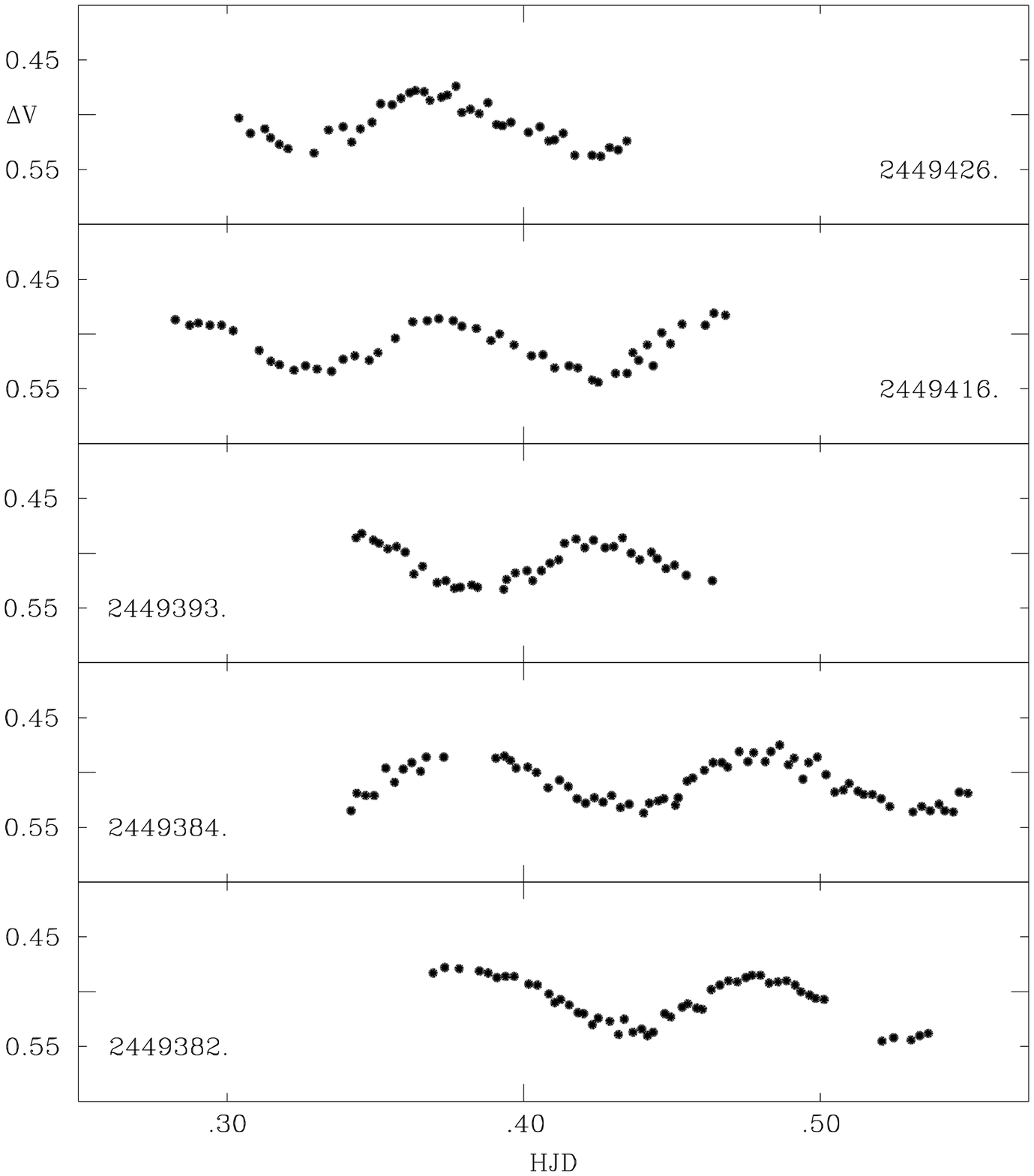} 
\caption[h]{Light curve of the $\delta$ Sct star AZ CMi as observed in
the same nights of Fig.1: the variability of the extinction
coefficient was corrected by our method and satisfactory light curves 
were obtained.}
\end{figure}

\section{The measure of the extinction coefficient}
When the measurements of the star brightness must be very accurate, it is
necessary to determinate the extinction coefficient $k_\lambda$ in a very
reliable way. In such a case, the usual Bouguer's line cannot be recommended
since the constancy of $k_\lambda$ (for the whole night in
all the directions) is a constraint very hard to satisfy. As we measured a lot
of times, slow variability of $k_\lambda$ is a common phenomenon in a 
night, also in sites considered as photometric ones, as the high altitude
ones (drifts of a few hundredths of mag/airmass are observed, see Poretti
\& Zerbi 1993a, 1993b). At lower altitude sites (Merate id 328 m above the
sea level) an
erratic and large amplitude (more than 0.10 mag/airmass) variability is 
more frequent. In general, it mimics the
variations in the lower part of the atmosphere, which is influenced by the 
local conditions on the ground: presence of woods, lakes, industrial and
urbanized areas ... Moreover, also temperature and relative humidity changes
can modify the $k_\lambda$ value, especially in the first hours of the night.
The passage of medium altitude clouds is also responsible for long term waves
in the measured flux. 

In such conditions, the differential photometry
is  strongly recommended to minimize these effects. However, it cannot 
ensure the requested accuracy when the fluctuations of $k_\lambda$ are large
even if the comparison stars are not very far. For this reason,
we developed a method allowing us to monitor continuously the transparence
variations. To do that, the instrumentation is kept under a tigth control
(cooled, always on, no hardware changes..); an artificial source is inserted
in the filter wheel to check at the beginning of each night if the instrumental
response is the same. We always verified that within the statistical errors,
confirming that a photon counting photometer can provide a high--level
stability.

Our main research field is the study of short period variable stars
(see Poretti \& Mantegazza 1992 and references therein); the 
target stars and its two comparison stars are measured for all the night.
In such a way, it is quite easy to build the graphs described
by Poretti \& Zerbi (1993a, 1993b) to calculate $m_0$, the {\it instrumental
magnitude above the atmosphere} of a comparison star. Once $m_0$ is known,
$k_\lambda$ can be evaluated every time this
star was measured, resolving the Bouguer formula
\begin{equation}
k_\lambda(t)=(m(t)-m_0)/X
\end{equation}
where $X$ is the airmass, $m(t)$ and $m_0$ are the magnitudes of the reference
star inside and outside the atmosphere, respectively. 
Figure 1 shows some examples of the variability
of the $k_V$ coefficient as observed when measuring the $\delta$ Sct star
AZ CMi at Merate Observatory (Poretti at al. 1996). As can be easily noticed,
it is hard to see a constant behaviour. In the JD 2449393 night the continuous
drift was of about 0.10 mag/airmass, while in the JD 2449384 night the
parabola--like
arc spans 0.15 mag/airmass. Note that fluctuations starting after the half 
of the JD 2449416 night, in a quite unpredictable way. Note also the presence
of waves in the $k_V$ curves, practically in all the nights, and the changing
mean level of the $k_V$ value, suggesting that an average value should not
exist.

The method described in the quoted
papers and briefly recalled here provides
a satisfactory correction  of the extinction since the instantaneous value
can be used all the times the magnitude difference stars are calculated.
Figure 2 shows the very small amplitude light curve of AZ CMi as obtained
from the measurements carried out in the same nights: as can be seen, the
regular cycle of variability is well described by the single points
(see Poretti et al. 1996 for further details on the frequency analysis).
It is important to note that its application does not
require any additional measurement, i.e. no time is subtracted to the
main programme.

\section{The sky brightness: the procedure to measure it}
 To measure the sky brightness, we used the
50--cm telescope and the photon--counting instrumentation. We give here a short
description of the procedure, emphasizing some methodological
aspects.

To establish
the relationship between counts per second and magnitude, a standard
star was measured and the sky as observed near it was subtracted. Therefore,
the absorption correction was applied to transform the catalogue magnitude
to the measured one. If the star is very bright,
the dead--time correction was also applied; Poretti (1992) describes a
reliable method to determine it.
Once we did that, the sky was measured
in different positions. To calculate the brightness per arcsec$^2$, we need
for the diaphragm area. The exact sizes of the diaphragms 
can be determinated
by the crossing time of a  bright star (an eyepiece allows us to see the 
diaphragm wheel); this procedure is suitable for larger ones, but
not recommended for smaller ones. The latter values can be better established
by performing some sky  measurements in the same position, allowing us to
obtain very accurate area ratios; once the larger areas were determined by
means of the crossing times, the smaller ones can be derived. In all these measures,
do not forget to subtract the value of the dark current.
\begin{figure}
\epsfysize=19truecm 
\epsfxsize=13truecm 
\hspace{3.5cm}\epsfbox{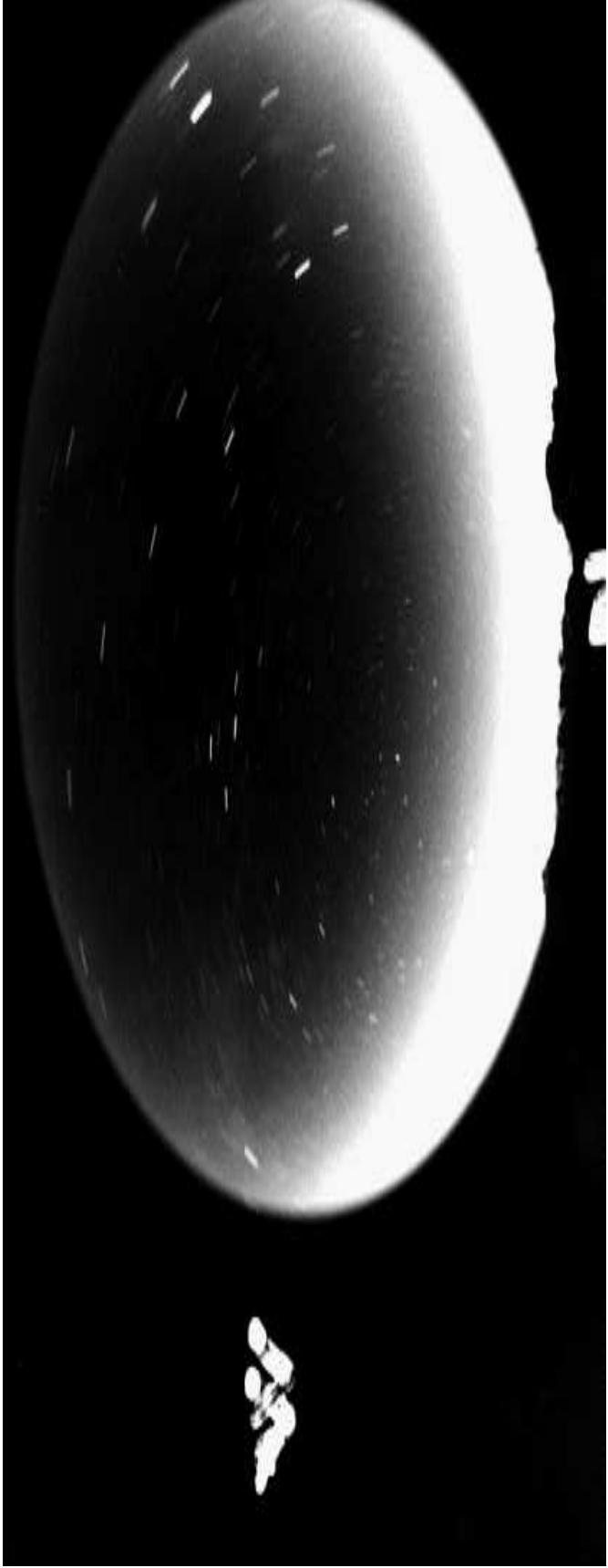} 
\caption[h]{Picture of the night sky taken with a fish--eye objective from
Merate Observatory. Exposure time was 15 minutes, November 1990 }
\end{figure}
\begin{figure}
\epsfysize=19truecm 
\epsfxsize=13truecm 
\hspace{3.5cm}\epsfbox{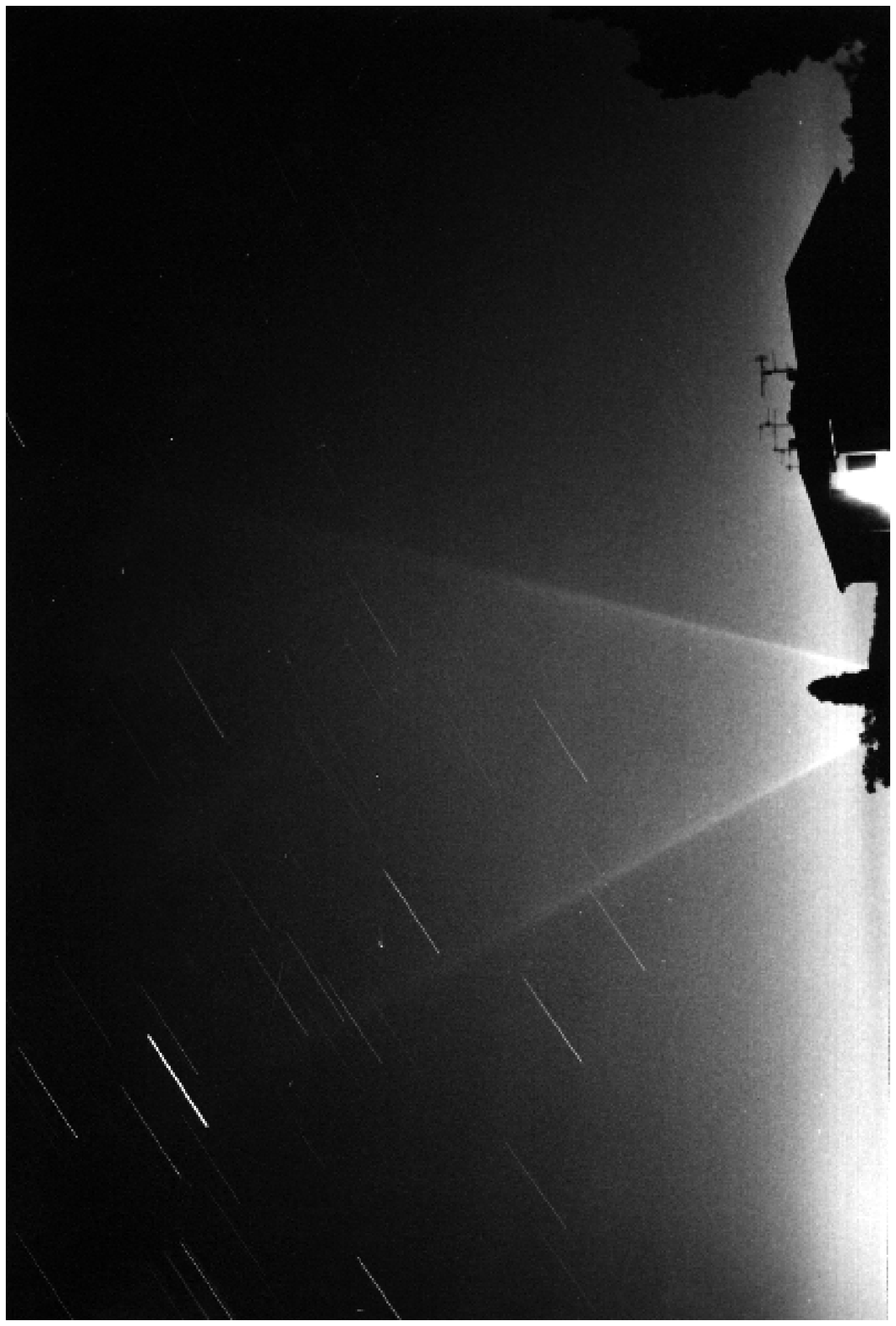} 
\caption[h]{The effect of a searchlight on the sky seen from Merate Observatory.
Its damage to the observational activities was demonstrated and its putting out
was decided by the public authority}
\end{figure}

\section{The sky brightness: the Merate values}
To check the dependence of the measured value from the position, we performed
a series of measurements in different parts of the sky  (September 1990).
 Of course, the darkest point was the zenith, where we measured 
$V$=18.2 mag/arcsec$^2$ in average local conditions ($k_V$=0.50 mag/airmass, 
no moon, no wind); this value was taken as a reference. 
Then we moved the telescope mapping the sky toward the south. Table~1 lists
the observed increases in brightness: note that 40--45$^o$ above the horizon
the sky is already 1.0 mag brighter than at zenith. To yield a better evidence
of this fact,  Fig. 3 shows a fish--eye picture
of the sky as seen from Merate observatory; the pollution by the light of
the neighbour towns (Merate, Milano, Lecco, Como,~...) is impressive.

\vspace{1cm} 
\begin{table}[h]
\centerline{\bf Tab. 1 - Increase of the sky brightness at different
declinations and hour angles values.}
\centerline{\bf (0 h, +45$^o$) corresponds to the
zenith value, i.e. $V$=18.2 mag/arcsec$^2$.}
\hspace{1.5cm} 
\begin{tabular}{|l c|l|l|l|}
\hline
 & &\multicolumn{3}{|c|}{Hour angle} \\
\multicolumn{1}{|c}{Decl.}& &\multicolumn{1}{|c|}{ -- 2 h} & 
\multicolumn{1}{|c|}{0 h} &\multicolumn{1}{|c|}{ +2 h}\\
\hline
+~45$^o$& &0.08 & 0.00 & 0.09 \\
+~35$^o$& &0.19 & 0.08 & 0.18 \\
+~25$^o$& &0.30 & 0.19 & 0.35 \\
+~15$^o$& &0.60 & 0.40 & 0.62 \\
+~05$^o$& &1.02 & 0.61 & 0.95 \\
--~05$^o$& &1.32 & 0.90 & 1.28 \\
\hline
\end{tabular}
\end{table}
\section{Preserving the observational capabilities}
We met a lot of problems in defending the capabilities of our 
observatory. Since it is located in a crowded area, there are often conflicting
interests from different people. However, we did an effort to create a 
favourable climate of opinion about our activity, especially by means of
straight contacts with local administrators. Since the results of the
measurements carried out in our Observatory are regularly published in
professional journals, this is a relatively easy task.

To avoid light dispersion toward the sky, we mainly try to convince them
to use cutoff lights. On the basis of a collaboration plan, they ask for our
approval about the lightings of large plants, even if the fulfilment of
our suggestions cannot be considered as mandatory. In general, we met an
attentive audience of our opinions. However, we stress the importance of
a regional or national law which establishes the safeguard of astronomical
sites, both from a scientific and a cultural point of view. In the
lack of that, disputes can easily arise. As  an example, we
had a legal controversial against a discotheque which placed a powerful
searchlight on the roof to be seen at large distances (Fig. 4).
On the basis of a collection of evidence, official institutions ordered to
switch off the searchlight since its damage to the observations carried out
in our institute was evident.



\acknowledgements
The authors wish to thank Sergio Cant\`u for the help in taking the sky
pictures. The generous help of the {\sc cariplo} Foundation in upgrading the
Ruths telescope is deeply acknowledged.


\end{document}